\documentclass[sigplan,10pt]{acmart}

\usepackage{pifont}
\usepackage{multirow}
\usepackage{makecell}
\usepackage[table]{xcolor}
\usepackage{enumitem}
\usepackage[ruled,linesnumbered]{algorithm2e}
\usepackage{subcaption}
\AtBeginDocument{%
  }

\setcopyright{none}
\copyrightyear{2026}
\acmYear{2026}
\acmDOI{}
\acmConference[ATC '26]{2026 ACM SIGOPS Annual Technical Conference}{November 16--18, 2026}{Shatin, Hong Kong}
\acmISBN{}
\renewcommand\footnotetextcopyrightpermission[1]{}
\AtBeginDocument{%
  \fancypagestyle{standardpagestyle}{%
    \fancyhf{}%
    \fancyfoot[C]{\footnotesize\thepage}%
  }%
  \fancypagestyle{firstpagestyle}{%
    \fancyhf{}%
    \fancyfoot[C]{\footnotesize\thepage}%
  }%
  \pagestyle{standardpagestyle}%
}

\begin{document}

% \title{KernelFlume: Achieving Fluid Inference via Kernel-Level Disaggregation and Asynchronous Pipelining}
\title[KernelFlume: Elastic Core-Attention Scaling for Agentic Long-Context Decoding]{KernelFlume: Elastic Core-Attention Scaling \\ for Agentic Long-Context Decoding}
%% Authors (arXiv public version) in ACM format.  The visible markers match
%% the previous compact version: # for equal contribution and * for
%% corresponding author.  Yuxin Wang is marked as having no affiliation.
\author{Guangyu Xiang\textsuperscript{\#}}
\affiliation{%
  \institution{HKUST(GZ)}
  \city{}
  \country{}}
\email{gxiang190@connect.hkust-gz.edu.cn}

\author{Xueze Kang\textsuperscript{\#}}
\affiliation{%
  \institution{HKUST(GZ)}
  \city{}
  \country{}}
\email{xkang507@connect.hkust-gz.edu.cn}

\author{Lin Zhang}
\affiliation{%
  \institution{HKUST}
  \city{}
  \country{}}
\email{lzhangbv@connect.ust.hk}

\author{Wenxiang Lin}
\affiliation{%
  \institution{HIT(SZ)}
  \city{}
  \country{}}
\email{wenxianglin@stu.hit.edu.cn}

\author{Shaohuai Shi}
\affiliation{%
  \institution{HIT(SZ)}
  \city{}
  \country{}}
\email{shaohuais@hit.edu.cn}

\author{Yuxin Wang\textsuperscript{*}}
\affiliation{%
  \institution{No affiliation}
  \city{}
  \country{}}
\email{yxwang.ele@gmail.com}

\author{Xiaowen Chu}
\affiliation{%
  \institution{HKUST(GZ)}
  \city{}
  \country{}}
\email{xwchu@hkust-gz.edu.cn}

\renewcommand{\shortauthors}{Xiang et al.}

\begin{abstract}
% LLM serving is increasingly dominated by long and dynamic decode workloads from agents, reasoning models, and extended conversations.
% When bursty long-context demand exceeds the capacity of a deployment, today's elasticity mechanisms add capacity at instance granularity by scaling out another model-serving replica.
% This expands KV memory only by duplicating the full model state, so elasticity inherits the startup cost and resource inefficiency of full-instance scaling.

LLM serving is increasingly dominated by long and dynamic decode workloads from agents, reasoning models, and extended conversations. When bursty long-context demand exceeds the deployed capacity, existing serving systems typically scale out by launching additional serving instances with model replicas. This instance-level elasticity increases KV capacity only by provisioning another full copy of the model, thereby inheriting the startup latency, memory overhead, and batch fragmentation.

We present KernelFlume, a decode-centric architecture that disaggregates the stable projection/FFN path from core-attention computation: weight nodes execute dense projection/FFN kernels, while weightless attention nodes store token-range KV partitions and scale with request-state demand.
To make this separation elastic, KernelFlume maintains a routing table that maps token ranges to attention-node endpoints.
It updates routes at token boundaries and uses host-visible graph signals to drive pre-registered UCX endpoint communication outside the captured CUDA Graph.
To preserve low per-token latency after disaggregation, KernelFlume combines query-first core-attention dispatch with inter-layer kernel pipelining, overlapping remote attention and communication with local projection/FFN work.
On real GPU testbeds (intra-node A6000 and cross-node H100), under a dynamic long-context agentic workload serving Llama-3.1-8B, KernelFlume sustains flat p99 TPOTs of ${\sim}$74\,ms on A6000 and ${\sim}$34\,ms on H100, while lowering cost per million output tokens by up to 32\% and 61\%, respectively, relative to full-instance elastic scaling with ServerlessLLM, a state-of-the-art instance-startup method.
Replaying the same trace at larger model scale in simulation projects a 56--66\% cost reduction over ServerlessLLM, widening to 80--85\% with cheaper heterogeneous attention-node hardware and persisting into the million-token context range.
\end{abstract}

%% CCS concepts and keywords (restored for the public arXiv version).




\maketitle

%% Unnumbered first-page footnote for the author-marker legend (#, *).
\newcommand{\blfootnote}[1]{%
  \begingroup
  \renewcommand\thefootnote{}\footnote{#1}%
  \addtocounter{footnote}{-1}%
  \endgroup
}
\blfootnote{$^{\#}$Equal contribution.\quad $^{*}$Corresponding author.}

\section{Introduction}

\begin{figure}[t]
    \centering
    \includegraphics[width=\linewidth]{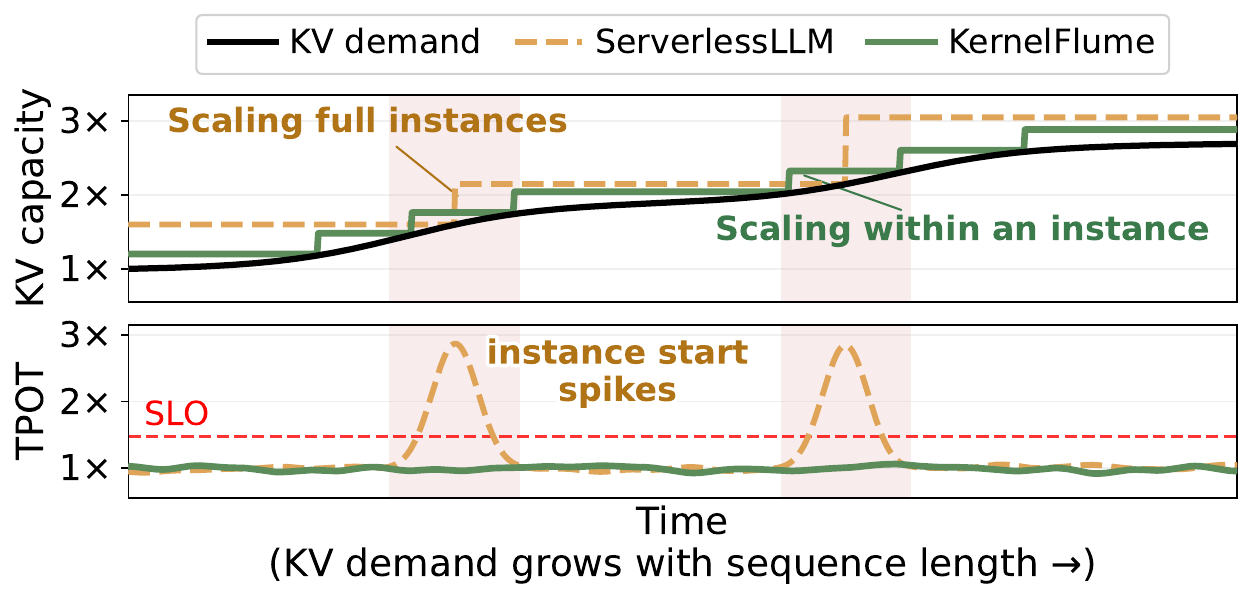}
    \caption{Conceptual KV-capacity scaling during long-context decode
    (schematic; measured results in \S\ref{sec:evaluation}).
    ServerlessLLM adds KV capacity only by scaling full instances; each
    scale-out (shaded) incurs an instance-start spike that pushes TPOT
    past the decode SLO. KernelFlume scales KV capacity \emph{within} the
    instance via weightless route updates that track demand, keeping
    TPOT flat.}
    \Description{Two stacked panels sharing a time axis along which KV
    demand grows. Top: relative KV capacity; a smooth rising demand curve, a
    chunky ServerlessLLM staircase with two large jumps, and a
    fine-grained KernelFlume staircase hugging the demand curve. Bottom:
    relative TPOT; both systems jitter around 1x in steady state, but
    ServerlessLLM spikes to about 2.9x at the two scale events, crossing
    the SLO line, while KernelFlume stays flat.}
    \label{fig:killfig}
\end{figure}

Large language models (LLMs) are increasingly deployed in decode-heavy, long-context workloads, including deep reasoning, autonomous agents, and multi-turn interactions~\cite{packer2023memgpt,wu2025dualpath,chen2025concur,guo2025deepseekr1,gao2024cachedattention,zheng2023lmsys,inferact2026codextraces}.
In these settings, the model does not simply process one large input and produce a short answer.
It generates tokens over many turns, carries intermediate state forward, and gradually builds a long live context.
LLM serving has two main phases: prefill processes the input context and builds the KV cache, while decode generates tokens one by one by reading and extending that cache.
When cross-turn KV-cache hits are high, repeated prefill work is largely reused; the remaining online work is sustained decode over a growing cache.
Recent Codex/SWE-bench Pro traces illustrate this pattern: agentic sessions carry a median context of roughly 80K tokens, with tails over 200K, and 94.2\% cross-turn KV-cache hits~\cite{qiao2026vllmmooncake,inferact2026codextraces}.

This shift makes KV cache capacity the central challenge of long-context decode. LLM decoding relies on storing keys and values for all previously processed tokens, so KV memory grows linearly with both batch size and sequence length, while model weights remain fixed. In these decode-heavy workloads, context expands continuously during decoding and quickly exhausts available high-bandwidth memory (HBM). For example, when serving Llama-3.1-8B (FP16) on an NVIDIA A6000 (48\,GB), a practical batch size of 64 leaves room for only about 4K tokens per request. Beyond this point, supporting longer sequences requires either reducing batch size, which hurts throughput, or provisioning more GPU memory capacity, typically by adding more GPUs.

However, in long-context decode, sequence length and KV growth are dynamic and unpredictable~\cite{xu2025memorywalls,wang2025burstgpt}.
Even when the serving stack smooths request arrivals, the relevant demand is not just the admitted request rate, but the accumulated KV state.
As a result, static provisioning either under-provisions dynamic KV bursts, delaying admission or reducing active decode concurrency, or over-provisions for peak demand, leaving expensive GPUs underused during common non-peak periods~\cite{yu2022orca,kwon2023vllm,agrawal2024sarathi,miao2024spotserve,fu2024serverlessllm}.
Efficient serving therefore requires elastic provisioning that adapts GPU capacity to changing KV demand.

Existing elastic serving methods typically use \textit{full-instance scaling}, adding or removing full instances, each carrying a complete model replica with the same parallelism configuration, as demand changes. As Figure~\ref{fig:killfig} illustrates, each scale-out incurs an instance-start spike that pushes time-per-output-token (TPOT) past the decode service-level objective (SLO). This approach remains inadequate for decode-heavy long-context workloads (\S\ref{sec:motivation}): instance starts are too slow for decode-time SLOs even with ServerlessLLM~\cite{fu2024serverlessllm}, a state-of-the-art instance-startup method; each scaling step carries a full weight copy that consumes HBM otherwise available for KV; and partitioning requests across instances shrinks the per-GPU batch and degrades utilization~\cite{liu2023ring,fu2024serverlessllm}.

Ultimately, these limitations highlight that the fundamental problem lies in the monolithic scaling unit. Because full-instance scaling intrinsically couples KV capacity with model weights, it forces a severe trade-off between elasticity, memory footprint, and compute efficiency. To expand KV capacity without replicating weights or fracturing the active batch, a serving system must scale KV capacity \emph{within} the serving instance by \emph{disaggregating attention-side capacity from model weights}, so that dense projection and FFN operations remain pinned to large, stable batches on dedicated GPUs, while only the KV cache dynamically scales up.

In this work, we present \emph{KernelFlume}, a decode-centric serving architecture that isolates projection and FFN on \emph{weight nodes}, while placing KV cache on scalable, \emph{weightless attention nodes}.

This disaggregation introduces two requirements. \textit{R1: fine-grained online elasticity.} Scale events must reconfigure KV capacity at token boundaries with minimal overhead. Route changes should not require model loading, communicator rebuilds, CUDA Graph recaptures, or pausing in-flight decode steps.
\textit{R2: SLO and cost efficiency.}
After KV capacity is scaled across attention nodes, KernelFlume must still meet strict TPOT SLOs and maintain low \$/Mtok.
The execution path should avoid duplicating weights or fragmenting the active batch, while overlapping remote attention with projection and FFN work to keep TPOT low.

This paper makes three contributions:
\begin{itemize}[topsep=3pt, leftmargin=*, noitemsep, nolistsep, parsep=0pt, partopsep=0pt]
  \item We identify the scaling mismatch between projection/FFN and attention during decode, and propose a disaggregated architecture that decouples KV capacity scaling from model-weight placement, enabling attention nodes to scale independently without replicating weights or shrinking the projection batch.
  \item We design an elastic routing control plane that manages attention-node membership through point-to-point routes, leaving the GPU execution graph unchanged across scale events. Once endpoints are prewarmed, activating a prepared route takes 7.2\,$\mu$s on the critical path, avoiding model reloads, communicator rebuilds, and disruption to in-flight decode steps.
  \item We introduce query-first attention disaggregation (QFA) and inter-layer kernel pipelining. Query-first dispatch allows $Q$ transfer to overlap with the remaining projection work, while the pipeline overlaps remote attention with weight-node projection and FFN work, keeping disaggregated decode within a few percent of the non-disaggregated reference.
\end{itemize}

Our \emph{measured} evaluation on a real intra-node NVIDIA RTX A6000 testbed and a cross-node H100 testbed demonstrates that under a real-world dynamic, long-context agentic workload from the Codex/SWE-bench Pro serving trace~\cite{qiao2026vllmmooncake,inferact2026codextraces}, KernelFlume maintains a flat, tight p99 TPOT (${\sim}$74\,ms on A6000 and ${\sim}$34\,ms on H100), with up to 32\% (A6000) and 61\% (H100) lower cost per million output tokens than full-instance elastic scaling with ServerlessLLM.
Beyond the measured testbeds, we replay the same trace at Llama-70B scale using measured hardware parameters, after validating the model against the measured cross-node 8B H100 testbed.
The simulation projects a 56--66\% cost reduction over ServerlessLLM at 70B scale, widening to 80--85\% with heterogeneous attention-node hardware, and preserves KernelFlume's cost advantage into the million-token context range (\S\ref{sec:eval-sim}).

\section{Background and Motivation}\label{sec:motivation}

\subsection{Agentic Long-Context Decoding}
\label{sec:long-dynamic}
LLM deployment is rapidly shifting from single-turn chatbots to decode-heavy long-context tasks such as deep reasoning and autonomous agentic workflows~\cite{guo2025deepseekr1,edgereasoning2025,xu2025memorywalls,wu2025dualpath,inferact2026codextraces}. In these workloads almost all GPU time goes to autoregressive decoding rather than prefill: on Codex/SWE-bench Pro traces, \textbf{94.2\%} of each turn's prompt tokens are served from the cross-turn KV cache~\cite{qiao2026vllmmooncake,inferact2026codextraces}, amortizing repeated prefill so that decode dominates end-to-end time. Decode efficiency therefore becomes the primary factor governing two key serving objectives: TPOT, which determines SLO compliance, and serving cost per million output tokens (\$/Mtok), which determines serving economy~\cite{miao2024spotserve,gao2024cachedattention}.

Achieving both goals is particularly difficult because these decode-heavy workloads share two distinguishing properties. First, they produce \ding{182} \emph{long contexts}: Codex/SWE-bench Pro agentic sessions carry a median context of roughly 80K tokens, with tails over 200K~\cite{qiao2026vllmmooncake,inferact2026codextraces}. Second, they exhibit \ding{183} \emph{high dynamism}. The growth rate varies unpredictably across requests and over time. Real-world traces show that request output lengths can span two orders of magnitude within the same workload, and bursty arrivals cause aggregate KV demand to swing sharply within minutes~\cite{wang2025burstgpt,xu2025memorywalls}. A serving system must therefore handle both large KV footprints and their rapid, unpredictable fluctuations.

\begin{figure}[t]
  \centering
  \includegraphics[width=0.98\linewidth]{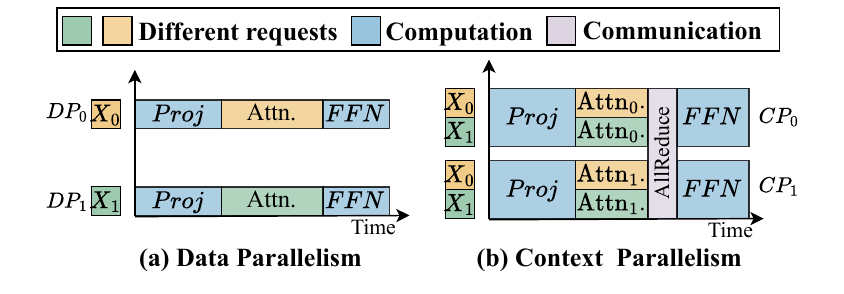}
  \caption{Two static scaling strategies for KV capacity. Data Parallelism (DP) replicates the full model and isolates requests, suiting large batches of shorter sequences. Context Parallelism (CP) partitions a single sequence's KV cache across GPUs, suiting long sequences.}
  \Description{A schematic comparing DP and CP execution for two requests, showing that projection and FFN are shared compute phases, while attention benefits from per-request partitioning and communication.}
  \label{fig:scaling-mismatch-a}
\end{figure}
\subsection{Scaling Long-Context Decoding}

Each autoregressive decode step (illustrated in Figure~\ref{fig:arch-inst}) performs three operations per layer: a \emph{QKV projection} that produces query ($Q$), key ($K$), and value ($V$) from the token embedding, an \emph{attention} kernel that reads the full KV cache to compute attention scores against $Q$, and a \emph{feed-forward network} (FFN) that yields the next-token logits. Projection and FFN are compute-bound over fixed model weights; attention is memory-bound and scales linearly with the accumulated KV cache length, making it the dominant bottleneck as sequences grow. As noted in \S\ref{sec:long-dynamic}, modern reasoning and agentic workloads routinely generate tens of thousands of tokens per session, quickly exhausting the KV capacity of a deployed GPU pool.

\textbf{Multi-GPU parallelism.} To accommodate this memory demand, serving systems organize multi-GPU deployments in two ways (Figure~\ref{fig:scaling-mismatch-a}). Data Parallelism (DP) creates multiple serving instances, while Context Parallelism (CP)~\cite{liu2023ring} places multiple GPUs inside one serving instance. In both strategies, serving ranks remain weight-bearing; the distinction is how requests and KV are mapped. Under DP, each request is assigned to one instance, and its KV cache remains within that instance. Under CP, an instance partitions each request's KV cache across its GPUs, allowing a sequence to span their aggregate memory. In short, DP splits the batch across independent serving instances, reducing per-instance KV pressure by shrinking the local batch, while CP extends per-sequence KV memory within a serving instance. Both, however, are static configurations fixed at deployment time.

\subsection{Need for Elasticity in Dynamic Decoding}
\begin{figure}[t]
  \centering
  \begin{subfigure}[t]{0.48\columnwidth}
    \centering
    \includegraphics[width=\linewidth]{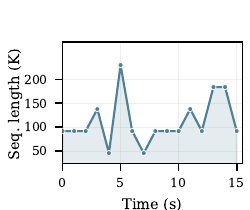}
    \caption{Sequence length over time.}
    \label{fig:intro-motivation-a}
  \end{subfigure}%
  \hfill
  \begin{subfigure}[t]{0.48\columnwidth}
    \centering
    \includegraphics[width=\linewidth]{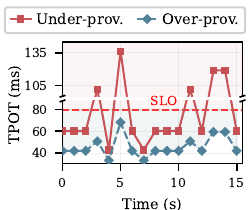}
    \caption{TPOT over time.}
    \label{fig:intro-motivation-b}
  \end{subfigure}
  \caption{Dynamic workloads make elasticity necessary. Replaying a dynamic long-context agentic workload~\cite{qiao2026vllmmooncake,inferact2026codextraces}, under-provisioning violates the SLO, whereas over-provisioning meets it but leaves excess capacity idle (TPOT sits far below the SLO), increasing \$/Mtok by 21.6\%.}
  \Description{A two-panel figure. Panel (a) shows per-batch context lengths from a dynamic long-context agentic workload (Codex/SWE-bench Pro trace). Panel (b) plots TPOT over the same window for two baselines: under-provisioning, which violates the SLO, and over-provisioning, which leaves excess capacity idle.}
  \label{fig:intro-motivation}
\end{figure}

The scaling strategies above assume a fixed workload, but the high dynamism identified in \S\ref{sec:long-dynamic} means real KV demand fluctuates sharply over time. Figure~\ref{fig:intro-motivation} makes this concrete by replaying a bursty long-context trace on a statically provisioned baseline. As shown in Figure~\ref{fig:intro-motivation-a}, recurring bursts of long-context requests cause aggregate KV demand to swing well above and below the provisioned capacity, exposing a fundamental provisioning trade-off.

In \emph{under-provisioning}, the system is provisioned for typical demand. When bursts arrive, KV capacity is exceeded, forcing the engine to queue or preempt requests and shrink the active batch. As Figure~\ref{fig:intro-motivation-b} shows, in our measurement this leads to only 68.8\% SLO attainment~\cite{kwon2023vllm}, with tail latency spiking during burst windows. In \emph{over-provisioning}, reserving enough capacity for the worst-case burst keeps TPOT within the SLO, but the extra GPUs sit idle during the far more common non-burst periods, increasing \$/Mtok by 21.6\%~\cite{miao2024spotserve,fu2024serverlessllm}. Neither regime is satisfactory, because both commit to a fixed amount of KV capacity at deployment time while the workload's KV demand is inherently dynamic. This motivates an \emph{elastic} serving architecture that can expand KV capacity when bursts arrive and release it when demand subsides, achieving both low TPOT and low \$/Mtok without a static provisioning compromise. The key question is the granularity of this elasticity: should the system scale an entire instance, or only the part of decode whose resource demand is growing?

\subsection{Limitations of Existing Elastic Scaling}

When aggregate KV demand exceeds the deployed pool, a serving system needs elasticity to add capacity. Scaling the two strategies above is constrained in different ways. 

\textbf{Scaling full instances.}
DP-style scale-out can add an instance and split the batch across instances, making expansion simple. This physically increases KV capacity, but only at the granularity of an instance, forcing KV capacity to scale together with compute capacity. The approach incurs three limitations for decode-heavy workloads. \emph{L1}, \emph{high instance-start overhead}. Bringing up a new instance requires loading model weights and initializing the serving runtime, and even optimized designs remain too slow for decode-time SLOs~\cite{fu2024serverlessllm}. \emph{L2}, \emph{low memory efficiency}. Each scaling step carries a complete set of model weights (e.g., 18\,GB for Llama-3.1-8B), consuming precious HBM that could otherwise hold KV cache. \emph{L3}, \emph{low compute efficiency}. Splitting the batch across instances shrinks each instance's local batch; in our measurement, halving the batch drops per-step decode throughput by 13--14\%, reducing GEMM utilization.

\textbf{Scaling within an instance.}
CP naturally matches sequence growth by partitioning KV across GPUs within an instance, but the participating GPU set is fixed at launch, so KV capacity cannot grow online with the context.
Fixed-pool sequence-parallel systems such as LoongServe~\cite{bai2024loongserve} improve the use of this predeployed pool through group reconfiguration, placement, and query dispatch, while avoiding resident-KV migration.
These mechanisms improve pool utilization but do not change its physical KV-capacity ceiling.
Expanding the pool online would require preparing a new weight-bearing rank, loading its model weights, changing the CP collective membership, and deciding how to handle resident KV.
The collective step alone is too slow for decode-time elasticity: rebuilding NCCL communicators for multi-GPU CP groups stalls execution for 550--662\,ms on our testbed.
Repartitioning resident KV is even more expensive: under high memory pressure on our testbed, a full-load resize can stall decode for seconds.
Thus, intra-instance scaling improves how a fixed pool absorbs long contexts, but adding KV capacity online beyond that pool is too expensive for decode-time elasticity.

Together, these two paths expose the missing scaling unit. Inter-instance scaling can add capacity online, but it expands KV capacity together with compute capacity. Intra-instance CP supports long sequences by partitioning each sequence's KV across GPUs, but only within a fixed pool. Long-context decode instead needs elasticity at the granularity of the growing KV state: expand KV-side capacity without copying model weights, changing the model-side execution path, or fragmenting the active batch. Turning this finer-grained scaling unit into a practical decode system imposes two requirements:

\emph{R1. Fine-grained online elasticity.} The system must reconfigure KV capacity at token-step granularity with minimal overhead. Route changes should not require model loading, communicator rebuilds, or pausing in-flight decode steps.
\emph{R2. SLO and cost efficiency.} After KV capacity is scaled across multiple nodes, the system must still meet strict TPOT SLOs and maintain low \$/Mtok.

\section{System Overview}
\label{sec:design}
% \begin{figure}[t]
%     \centering
%     \includegraphics[width=\linewidth]{Figures/overview/arch_v1.pdf}
%     \caption{KernelFlume's architecture comprises weight nodes and an elastic set of attention nodes. (a) Full-instance scaling adds a complete weight-bearing instance. (b) KernelFlume adds a weightless attention node and extends the token-range routing table.}
%     \Description{A system architecture diagram contrasting monolithic full-instance scaling with KernelFlume. The monolithic path adds a complete instance, while KernelFlume keeps model weights on weight nodes, adds a weightless attention node with a new KV range, updates the routing table, and executes a five-step decode flow.}
%     \label{fig:arch}
%     \end{figure}

\begin{figure}[t]
    \centering
    \begin{subfigure}[t]{\linewidth}
        \centering
        \includegraphics[width=\linewidth]{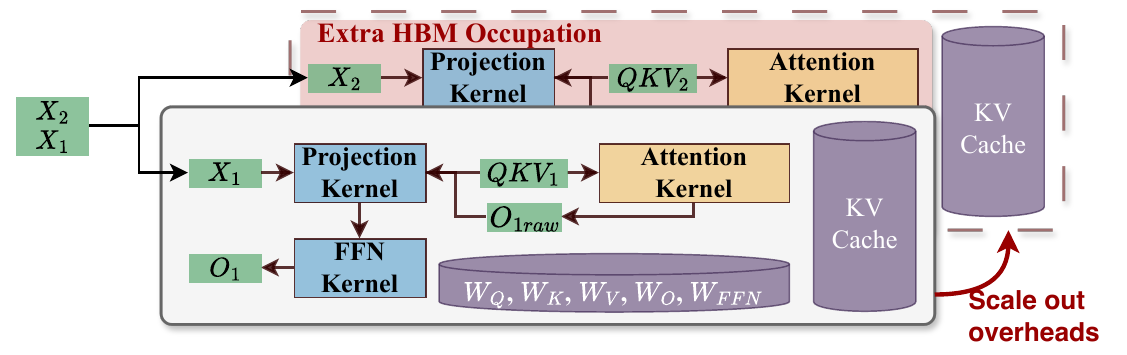}
        \caption{Instance-level elasticity for context growth (e.g., ServerlessLLM~\cite{fu2024serverlessllm})}
        \label{fig:arch-inst}
    \end{subfigure}
    \vspace{0.4em}
    \begin{subfigure}[t]{\linewidth}
        \centering
        \includegraphics[width=\linewidth]{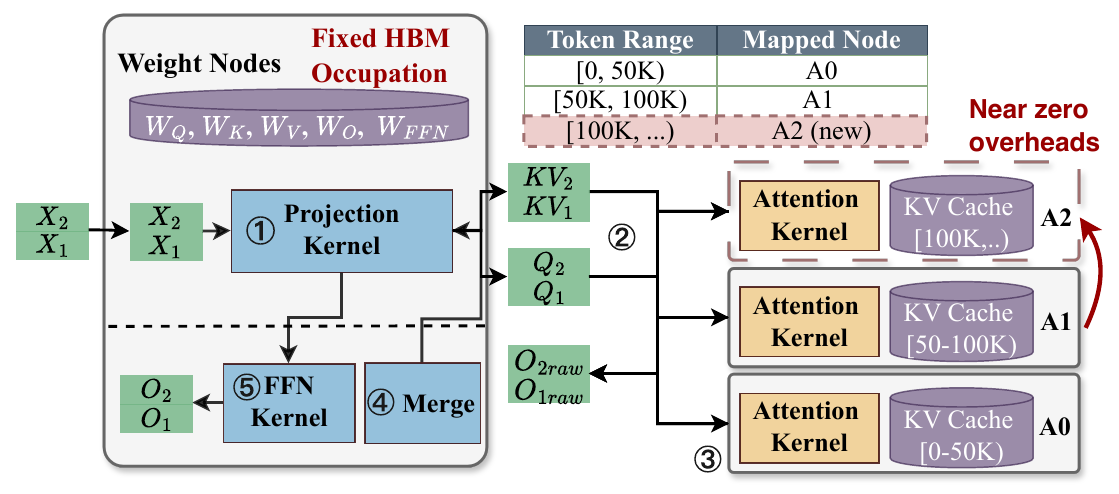}
        \caption{Architecture of KernelFlume}
        \label{fig:arch-kernelflume}
    \end{subfigure}
    \caption{KernelFlume's architecture comprises weight nodes and an elastic set of attention nodes. (a) Full-instance scaling adds a complete weight-bearing instance. (b) KernelFlume adds a weightless attention node and extends the token-range routing table.}
    \Description{Two vertically stacked architecture diagrams. The first shows instance-level elasticity for context growth, while the second shows attention-level elasticity that adds attention-side capacity for KV growth.}
    \label{fig:arch}
    \end{figure}

Prefill/decode disaggregation~\cite{zhong2024distserve,patel2024splitwise,qin2024mooncake} has shown that separating inference phases onto specialized hardware improves utilization. KernelFlume extends this philosophy to a finer granularity, \emph{within the decode phase itself}, separating projection/FFN from attention so that the attention/KV path scales elastically with growing request state.

Concretely, KernelFlume places weight (W) nodes alongside an elastic pool of weightless attention (A) nodes. Figure~\ref{fig:arch} contrasts the scaling unit: full-instance scaling adds another weight-bearing replica, whereas KernelFlume adds only a weightless A~node and extends the routing table.
KernelFlume addresses \textbf{R1} with this routing-table interface: routed communication sends tensors according to the table (\S\ref{sec:elastic-attn-comm}), and the scaling policy decides when to prepare and publish A~nodes (\S\ref{sec:scaling-policy}).
It addresses \textbf{R2} with QFA, which overlaps $Q$ transfer with remaining projection work (\S\ref{sec:qfa}), and inter-layer kernel pipelining, which overlaps A-side attention with W-side work across adjacent microbatches (\S\ref{sec:microbatch-pipeline}).
This section gives the architectural map and a decode-step walkthrough.

\subsection{Kernel-level Disaggregation}

KernelFlume's base architecture uses two node types, illustrated in Figure~\ref{fig:arch-kernelflume}. A \emph{weight node} is the weight-bearing deployment unit: a single GPU when the model fits on one device, or a tensor-parallel group for larger models. The elastic dimension is implemented by \emph{weightless} attention nodes, so ranks added for attention/KV capacity do not carry model weights.
A \textbf{weight (W) node} owns the model-weight path during decode, executing $W_Q$, $W_K$, $W_V$, merging attention partials, applying $W_O$, and running FFN locally.
Keeping these kernels on weight nodes preserves large projection/FFN batches and avoids weight movement during elasticity events.
An \textbf{attention (A) node} is a weightless worker that stores a token-range KV partition and computes local core attention over that partition.
Adding one changes only KV ownership and routing, so KernelFlume uses sequence partitioning as an attention-only scaling mechanism rather than a full-model parallel execution strategy.

This architecture aligns the scaling unit with the resource asymmetry identified in \S\ref{sec:motivation}: model weights stay on W~nodes, while A~nodes scale with the context-dependent attention/KV demand.

\subsection{Workflow of Elastic Decoding }

Figure~\ref{fig:arch-kernelflume} also summarizes the layer-level decode flow.
For each request, the routing table maps token ranges to the A~nodes that store their KV cache (\S\ref{sec:elastic-attn-comm}).
With this mapping, each decode layer proceeds as follows.

\emph{\ding{182} Projection on weight nodes.} Weight nodes run QKV projection for the current decode step. \emph{\ding{183} Routed dispatch by QFA.} With QFA (\S\ref{sec:qfa}), weight nodes dispatch $Q$ through the routed communication path (\S\ref{sec:elastic-attn-comm}) to all A~nodes that hold KV ranges for the request, as soon as $Q$ is ready and before the full QKV projection completes. The newly generated $(K_{\text{new}}, V_{\text{new}})$ is routed only to the request's \emph{tail attention node}, which owns the newest token range. Tail nodes change only at step boundaries. \emph{\ding{184} Local attention on the attention nodes.} After QFA dispatch, A~nodes that own only historical ranges for a request can begin attention as soon as $Q$ arrives. A~nodes that own live tail ranges wait for $(K_{\text{new}}, V_{\text{new}})$, append them, and then compute attention on the updated tail partitions. \emph{\ding{185} Partial-result return and merge.} Each attention node returns its local attention partials $(o_i, m_i, \ell_i)$ to weight nodes, which combine them with the standard online-softmax (log-sum-exp) reduction~\cite{milakov2018online,dao2022flashattention,liu2023ring} and apply $W_O$. \emph{\ding{186} FFN and next layer.} Weight nodes run FFN locally and advance to the next layer; inter-layer kernel pipelining (\S\ref{sec:microbatch-pipeline}) schedules this post-attention work against projection and remote attention from adjacent microbatches.

Scale events are applied between decode steps: the scaling policy prepares new A~nodes in the background and publishes route updates only at token boundaries (\S\ref{sec:scaling-policy}). An in-flight token therefore uses one consistent routing snapshot, while future tokens can append KV to newly added A~nodes. The next two sections detail the elasticity mechanisms and execution optimizations, respectively.

\section{Elastic Core Attention on Static Graphs}
\label{sec:dynamism}

\textbf{R1} requires scale events to take effect between decode steps without draining in-flight requests or rebuilding the execution path.
In KernelFlume, core-attention elasticity is implemented as dynamic routing: token ranges may map to different A~node endpoints, while the W~node execution graph remains unchanged.
This design must address three questions.
\ding{182} How can attention communication change its endpoints without rebuilding a collective group?
\ding{183} How can dynamic routes be exercised from a static CUDA Graph without graph recapture or serializing remote attention?
\ding{184} How can KernelFlume decide when to prepare and activate new attention nodes so that growing KV demand is absorbed without violating the TPOT SLO?
KernelFlume answers the first two questions with elastic attention communication (\S\ref{sec:elastic-attn-comm}), which separates endpoint selection from graph execution, and answers the third with a scaling policy (\S\ref{sec:scaling-policy}).

\subsection{Elastic Communication}
\label{sec:elastic-attn-comm}

\textbf{Routing table.}
Collective communication with a fixed peer set, such as an NCCL communicator~\cite{nvidia2023nccl} or an NVSHMEM processing-element group~\cite{nvshmem}, exposes the wrong granularity for online attention-node elasticity.
With collective-based communication, changing that peer set typically requires communicator rebuild or heap re-creation, participant rebinding, and draining in-flight work before the new topology becomes usable.
Those steps are appropriate for static tensor-parallel deployments, but they are too heavy for the sub-millisecond route switches that KernelFlume targets on the request-state path: on our testbed, rebuilding NCCL communicators alone takes 550--662\,ms.
The mismatch is that collective membership is part of the communication object: a newly added attention node cannot enter the decode path without rebuilding that object.

KernelFlume instead defines attention communication around a routing table rather than a collective group.
The routing table maps logical token ranges to UCX endpoint handles~\cite{shamis2015ucx}, along with metadata such as the active attention-node set.
For each request, it identifies the endpoints that own historical KV ranges and the endpoint that owns the live tail.
Weight nodes then communicate directly with those endpoints: they dispatch $Q$ to the owners of the request's token ranges, send $(K_{\text{new}},V_{\text{new}})$ only to the tail endpoint, and gather local attention partials from the same set.
Under batching, the transport-visible active set is the union of per-request endpoint sets.
A scale event edits entries for future KV ranges and updates the derived endpoint set, without changing any collective execution group.
This is the only decode-path state it changes: future token ranges are assigned to the new endpoint, while graph-captured computation remains unchanged.

UCX implements the data movement and lets the same routed endpoint abstraction use the best available GPU-aware transport for each endpoint pair.
The CPU only initiates transfers and never touches payloads, so the data plane runs at transport-native bandwidth.
Receives are pre-posted at the beginning of each decode step into GPU buffers consumed by the receiver's graph.
This zero-copy, pre-posted design eliminates staging copies and protocol-level rendezvous stalls: when a send arrives, the transport can begin DMA immediately without waiting for a matching receive.

\begin{figure}[t]
  \centering
  \includegraphics[width=\linewidth]{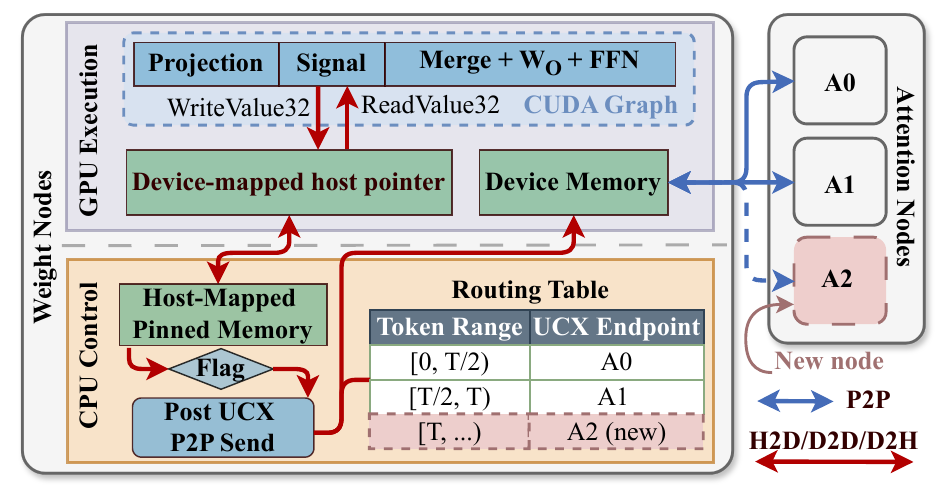}
  \caption{Elastic routing on a static graph. The topology-unaware weight-node CUDA Graph (top) contains local compute and host-visible readiness signals. The CPU-side routing loop (bottom) reads those signals and uses the routing table that maps token ranges to UCX endpoints. In this example, a request's KV cache initially spans A0~$[0, T/2)$ and A1~$[T/2, T)$. When the context grows beyond $T$, the routing table is extended with a new entry $[T, \ldots) \to \text{A2}$ without modifying the weight-node execution graph, turning scale-up into a routing-table update.}
  \Description{A two-layer diagram of weight nodes. The GPU execution layer contains a fixed CUDA Graph with projection, signaling, merge, and FFN stages. The CPU routing layer contains a routing table mapping token ranges to attention-node UCX endpoints, with entries for A0, A1, and a newly added A2. Three attention nodes are shown on the right, with A2 highlighted as the new node.}
  \label{fig:elastic-topo}
  \end{figure}

\textbf{Graph progress signaling.}
The routing table determines where tensors should be sent; the remaining question is how to issue those sends as soon as graph-produced buffers are ready.
CUDA Graph replay creates two constraints.
Encoding communication targets inside the graph would bind it to a fixed attention-node set and require recapture after scale events, while waiting until graph completion would serialize communication with computation and eliminate query-first overlap (\S\ref{sec:qfa}).
KernelFlume therefore separates graph execution from communication decisions.
Figure~\ref{fig:elastic-topo} illustrates the two paths: the upper path is a topology-unaware weight-node CUDA Graph that produces tensors and writes host-visible readiness signals, while the lower path is a CPU-side routing loop that reads those signals, consults the current routing table, and posts UCX sends to the selected endpoints.
A scale-up event updates the routing table, not the weight-node execution graph.

\textit{GPU path.}
The weight-node CUDA Graph contains only local compute and signal writes.
KernelFlume inserts signal points where outbound buffers become ready, such as after the $Q$ projection and after the $K/V$ projection.
Each signal uses a \texttt{WriteValue32}-style write to set a flag in host-visible memory, allowing the CPU to observe graph progress without ending replay or embedding communication targets in the graph.
Because token ranges and attention-node membership are absent from the graph, route changes do not require graph recapture.
\textit{CPU path.}
A CPU polling loop runs alongside the graph.
When a readiness flag is set, it consults the current routing table and posts UCX sends for the ready buffer to the selected A-node endpoints.

After all receives for a layer complete, the CPU writes a completion flag to GPU memory so the receiver-side graph's \texttt{ReadValue32}-style wait can resume.
This write uses a \emph{dedicated signaling stream} rather than the graph's execution stream: the execution stream is already blocked on that wait, so enqueuing the flag write on the same stream would place it behind the wait and deadlock.

\textbf{Transport pre-registration.}
KernelFlume exposes a route only after its transport state is ready.
Before an A~node receives routed KV ranges, KernelFlume creates its UCX endpoints and registers the GPU buffers on the routed path.
This moves CUDA IPC setup and GPUDirect RDMA registration outside graph replay, leaving decode steps to post sends to ready endpoints.
The same rule applies during scale-up: KernelFlume prewarms the new A~node and publishes it in the routing table only after these registrations finish (\S\ref{sec:scaling-policy}).

Together, endpoint routing, graph-external signaling, and pre-registered routes keep scale events outside the captured graph.
A route update changes only CPU-side destination metadata; tensor shapes and the W~node graph remain unchanged.
The CPU path is lightweight in practice: each routing-loop operation costs only a few microseconds in our measurements (\S\ref{sec:eval-controller}).
Because the loop runs alongside CUDA Graph replay and reacts to readiness signals, communication is issued when individual buffers become ready rather than after graph completion.

\subsection{Elastic Scaling Policy}
\label{sec:scaling-policy}

Given elastic communication and static-graph support, the remaining policy question is when to prewarm a new A~node, including its endpoint and buffer registrations, and when to publish its route in the routing table.

\begin{algorithm}[t]
  \caption{A-node scaling policy}
  \label{alg:scaling-policy}
  \KwIn{batch $\{s_1, \dots, s_B\}$; model constants $H_{kv}, d_h, L$; policy constants $\beta_{\text{mem}}, \beta_{\text{slo}}, \gamma$; SLO-derived $T_A^{\max}$; profiled $BW_{\text{HBM}}, C_{\text{kv\_cap}}, T_{\text{step}}, T_{\text{prewarm}}$}

  $C_{\text{thresh}} \leftarrow \min(\beta_{\text{mem}} C_{\text{kv\_cap}},\; \beta_{\text{slo}} T_A^{\max} BW_{\text{HBM}})$\;
  $C_{\text{total}} \leftarrow \sum_i 2 s_i H_{kv} d_h L \cdot 2\text{B}$\;
  $N_A \leftarrow \lceil C_{\text{total}} / C_{\text{thresh}} \rceil$\;
  Partition token ranges to balance aggregate KV bytes across the batch\;
  Initialize $\mathrm{tail}[r]$ for each request $r$\;

  \ForEach{tail-holding A-node $i$}{
    $n_i \leftarrow |\{r: \mathrm{tail}[r]=i \textbf{ and } r \text{ active}\}|$\;
    $g_i \leftarrow n_i \cdot 2 H_{kv} d_h L \cdot 2\text{B}$\;
    $b_i \leftarrow \text{KVBytes}(A_i)$\;
    \If{$g_i > 0$}{
      $\tau_{\text{fill}}^{(i)} \leftarrow (C_{\text{thresh}} - b_i) / g_i$\;
      \If{$\tau_{\text{fill}}^{(i)} T_{\text{step}} < T_{\text{prewarm}}$ \textbf{and} no pending prewarm for $i$}{
        $\mathrm{prewarm}[i] \leftarrow$ spawn prewarm task\;
      }
    }
    \If{$b_i \geq \gamma C_{\text{thresh}}$ \textbf{and} $\mathrm{prewarm}[i]$ ready}{
      At the next token boundary, publish the route and redirect active tails on $i$ to $\mathrm{prewarm}[i]$\;
    }
  }
  \end{algorithm}
\textbf{Per-node budget.}
The scaling policy provisions A~nodes against two constraints: KV capacity and the TPOT SLO.
HBM capacity gives a hard memory limit.
The SLO gives a latency limit: even when the KV state fits, the predicted exposed decode time must remain below the TPOT target.
Query-first dispatch reduces layer-local exposure by sending $Q$ before the full projection finishes (\S\ref{sec:qfa}), while kernel pipelining overlaps remaining attention-side time with W~node projection and post-attention work (\S\ref{sec:microbatch-pipeline}).
The pipeline model in \S\ref{sec:microbatch-pipeline} predicts the exposed per-layer time as $\max(T_p+T_f,\;T_A)$, where $T_p+T_f$ is the W-side stage and $T_A$ is the routed attention stage.
Given a TPOT target $T_{\text{slo}}$, KernelFlume subtracts measured per-token overheads outside this two-stage model and divides the remaining budget across layers, yielding a per-layer budget $T_{\text{layer}}^{\text{bud}}$.
For the chosen W-side configuration, satisfying the SLO requires $\max(T_p+T_f,\;T_A)\leq T_{\text{layer}}^{\text{bud}}$, so the largest admissible routed-attention time is $T_A^{\max}=T_{\text{layer}}^{\text{bud}}$.
KernelFlume then converts this time bound into a KV-byte budget for each A~node.
Because decode attention is bandwidth-bound, scanning a KV footprint of $C$ bytes takes approximately $C/BW_{\text{HBM}}$; the SLO-derived budget therefore gives $C_{\text{slo}} = \beta_{\text{slo}} T_A^{\max} BW_{\text{HBM}}$, where $BW_{\text{HBM}}$ is the effective HBM bandwidth of one weightless A~node.
The factor $\beta_{\text{slo}}<1$ reserves headroom for profiling error, kernel variability, and communication jitter.
The memory budget is $C_{\text{mem}} = \beta_{\text{mem}} C_{\text{kv\_cap}}$, where $C_{\text{kv\_cap}}$ is the usable KV capacity of one A~node.
The factor $\beta_{\text{mem}}<1$ leaves space for allocator fragmentation, runtime buffers, and transient non-KV state.
The policy therefore caps each A~node at
\begin{equation}
    C_{\text{thresh}} = \min(\beta_{\text{mem}} C_{\text{kv\_cap}},\; \beta_{\text{slo}} T_A^{\max} BW_{\text{HBM}}),
\end{equation}
where $C_{\text{thresh}}$ is the effective per-node threshold used by the policy, i.e., $\min(C_{\text{mem}}, C_{\text{slo}})$.
The tighter of the memory and latency bounds determines when another A~node is needed.

\textbf{Post-prefill initialization.}
After prefill, the batch has total KV footprint $C_{\text{total}} = \sum_i 2 s_i H_{kv} d_h L \cdot 2\text{B}$, where $s_i$ is request $i$'s token length, $H_{kv}$ the number of KV heads, $d_h$ the per-head dimension, $L$ the number of layers, and $2\text{B}$ FP16 storage per key or value element.
The policy provisions $N_A = \left\lceil C_{\text{total}} / C_{\text{thresh}} \right\rceil$ A~nodes and initializes the routing table by assigning existing token ranges to balance aggregate KV bytes across them.
For each request, the node that owns its highest token range becomes the current tail.

\textbf{Predictive scale-up.}
During decode, only tail nodes grow.
Non-tail nodes hold fixed historical ranges, while each active request whose tail is on node $i$ appends $2 H_{kv} d_h L \cdot 2\text{B}$ bytes per step.
For a tail-holding node $i$ with $n_i>0$ active requests and current KV occupancy $b_i$, KernelFlume estimates the fill horizon under the current growth rate as
\begin{equation}
    \tau_{\text{fill}}^{(i)} = \frac{C_{\text{thresh}} - b_i}{n_i \cdot 2 H_{kv} d_h L \cdot 2\text{B}},
\end{equation}
and recomputes this estimate every step from exact KV occupancy and tail ownership.
Here $T_{\text{step}}$ is the predicted decode-step time, $T_{\text{prewarm}}$ is the measured time to prepare a new route, and $\gamma$ is the activation watermark.
When $\tau_{\text{fill}}^{(i)}T_{\text{step}} < T_{\text{prewarm}}$, KernelFlume prewarms a fresh A~node in the background.
After the watermark is reached and the prewarmed node is ready, KernelFlume publishes the prepared route at the next token boundary.
Future appends use the new A~node, while existing KV ranges remain in place.
The critical-path route installation measured in \S\ref{sec:eval-controller} takes 7.2\,$\mu$s, with all endpoint setup performed off the critical path during background pre-warming.
Algorithm~\ref{alg:scaling-policy} summarizes the post-prefill initialization and the per-step monitor.

\section{Disaggregated Kernel Execution}
\label{sec:pipeline}

This section addresses \textbf{R2} by showing how KernelFlume sustains strict TPOT SLOs and low \$/Mtok when decode execution spans multiple nodes. It targets two sources of underutilization: exposed communication within each layer, which query-first attention reduces by overlapping communication with computation, and staggered work across layers, which inter-layer kernel pipelining reduces by overlapping adjacent microbatches.

\subsection{Query-First Attention}
\label{sec:qfa}

Query-first attention exposes and exploits the fact that many decode-time attention partitions depend only on the new query.
For a decode token, an attention partition that owns only historical KV already has its $K/V$ inputs; its only new dependency is $Q$.
In a non-disaggregated layer, this $Q$-only dependency is hidden by the usual kernel boundary: attention starts after the full QKV projection has produced all $Q$, $K$, and $V$ outputs.
KernelFlume implements this idea with \emph{query-first attention disaggregation}.
As soon as $Q$ is produced, KernelFlume routes it to the A~nodes that already hold historical KV, allowing their remote attention work to overlap with the remaining $K/V$ projection and other local work on W~nodes.

\begin{figure}[t]
    \centering
    \includegraphics[width=\linewidth]{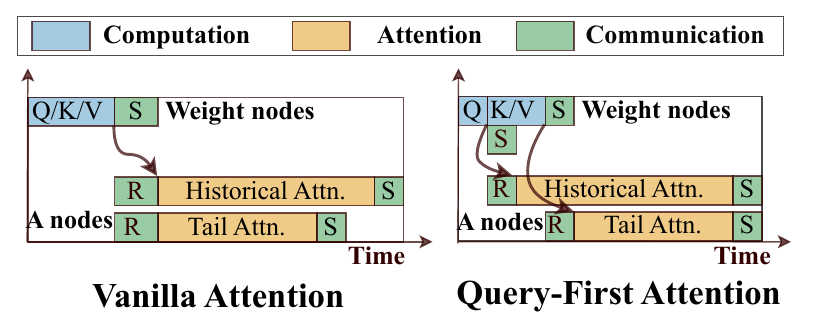}
    \caption{Layer-level timelines for normal attention (left) and query-first attention (right). With query-first attention, $Q$ is dispatched before the full QKV projection completes, allowing attention nodes to begin computing on historical KV in parallel with weight nodes' remaining K/V projection.}
    \Description{Two layer-level timelines. The top timeline shows serialized decode, where full projection completes before any remote attention work begins. The bottom timeline shows KernelFlume overlap, where $Q$ is dispatched early so non-tail attention nodes can start on historical KV while the weight nodes continue the remaining projection work.}
    \label{fig:timeline}
\end{figure}

Figure~\ref{fig:timeline} illustrates the resulting overlap.
Once $Q$ is dispatched, non-tail A~nodes can begin attention over their historical KV partitions while W~nodes continue $K/V$ projection and other local work.
For tail A~nodes, the current token also extends the local KV range: they wait for both $Q$ and $(K_{\text{new}},V_{\text{new}})$, append the new entries, and then compute attention on the updated tail partitions.
Each attention node returns local partials $(o_i,m_i,\ell_i)$, which the weight nodes merge via the associative online-softmax reduction~\cite{milakov2018online,dao2022flashattention} before applying $W_O$ and FFN.
This merge produces the same result as computing attention over the full KV range, so query-first dispatch and the later tail-node appends of $(K_{\text{new}},V_{\text{new}})$ preserve the kernel semantics.
Starting historical-KV attention as soon as $Q$ is ready lets A~nodes work during the remaining $K/V$ projection window, making added attention capacity effective within the layer.

The challenge is to expose $Q$ early while preserving CUDA Graph replay on W~nodes.
W~nodes run projection as a CUDA Graph for low-overhead replay, but once launched the graph is a fixed GPU command sequence: the CPU cannot interrupt it after $Q$ to issue a routed send before $K/V$ completes.
Splitting the projection into multiple graph launches would add launch and synchronization overhead, while waiting for graph completion would serialize communication and lose query-first overlap.

KernelFlume keeps the projection path in a single CUDA Graph and moves endpoint selection and send issuance to a CPU-side loop.
The graph writes $Q$ and $(K_{\text{new}},V_{\text{new}})$ into pre-registered GPU buffers and emits a host-visible readiness flag when each buffer is complete.
A dedicated CPU polling thread observes these flags, consults the current routing table, and posts non-blocking UCX sends to the selected A~nodes.
Thus, $Q$ can be in flight while the same graph launch continues the remaining $K/V$ projection.
On the receiver side, each A~node launches attention only after transport completion marks the input buffers ready, ensuring that kernels consume complete tensors.

This split preserves the invariants needed for query-first execution.
The W~node graph remains a deterministic, fixed-shape projection path and retains the efficiency of CUDA Graph replay.
At the same time, endpoint choices stay outside the graph: route updates described in \S\ref{sec:dynamism} change only the CPU-side routing table and do not trigger graph recapture.

\subsection{Kernel Pipelining}
\label{sec:microbatch-pipeline}
The query-first kernel schedule reduces layer-local communication exposure. Across layers, however, W~nodes can still wait for remote attention partials before they can merge the result, apply $W_O$, and execute FFN for the same microbatch.
KernelFlume addresses this with an \emph{inter-layer kernel pipeline}.

\begin{figure}[t]
    \centering
    \includegraphics[width=\linewidth]{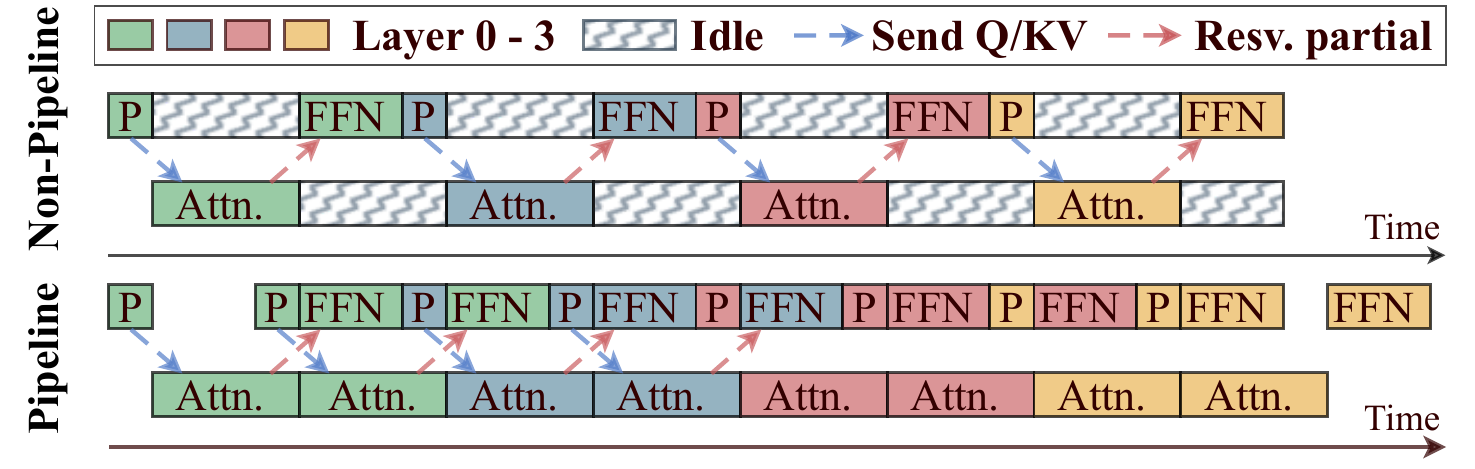}
    \caption{Serialized (top) vs.\ pipelined (bottom) execution with $M{=}2$ microbatches across four layers. Pipelining fills weight nodes' idle gaps by interleaving projection and FFN from alternating microbatches.}
    \Description{A two-part execution timeline for the kernel pipeline. The top part shows serialized execution with visible idle gaps while remote attention runs. The bottom part shows a pipelined schedule across four layers and two microbatches, where projection and FFN on the weight nodes overlap with attention on the attention nodes. Colored blocks denote layers and dashed arrows denote cross-node communication.}
    \label{fig:pipeline}
\end{figure}

Rather than processing all active decode requests as a single decode batch, KernelFlume groups the current decode traffic into $M$ \emph{microbatches} with target size $B_\mu$.
Write the $k$th microbatch as $\mu_k$.
For each layer, W~nodes execute $\texttt{proj}(\mu_k, L)$ and dispatch $Q$ to the A~nodes.
While A~nodes compute attention for $\mu_k$, W~nodes use the wait time to complete the merge, $W_O$, and FFN work for the previous microbatch $\mu_{k-1}$ whose attention result has already returned.
Once the pipeline is filled, W~nodes alternate between projection and post-attention work across microbatches, while A~nodes process the corresponding attention work.
Figure~\ref{fig:pipeline} illustrates the effect with $M{=}2$. The serialized schedule (top) leaves W~nodes idle whenever they wait for remote attention, while the pipelined schedule (bottom) fills those gaps, keeping both W and A~nodes close to continuously busy.

For the chosen microbatch size $B_\mu$, let $T_p$ be the profiled W-node projection time, $T_f$ be the profiled W-node post-attention time including merge, $W_O$, and FFN, and let $T_A(N_A)$ be the predicted A-side attention-stage time with $N_A$ active A~nodes, including attention compute and communication.
In a simplified two-stage model, serial execution spends $T_p+T_A(N_A)+T_f$ per layer for one microbatch, while the filled pipeline completes one microbatch every $\max(T_p+T_f,\;T_A(N_A))$.
The resulting ratio of pipelined throughput to serial throughput is
\[
    G_{\text{pipe}} =
    \frac{T_p + T_A(N_A) + T_f}
         {\max(T_p + T_f,\; T_A(N_A))}.
\]
The ideal upper bound is $2\times$ when the W-side stage and the A-side attention stage take similar time.
The same model gives the exposed per-layer time, $\max(T_p + T_f,\; T_A(N_A))$, which KernelFlume uses to predict decode-step latency for SLO-aware scaling.
As sequences grow, the scaling policy in \S\ref{sec:scaling-policy} uses this estimate to add A~nodes before the attention stage threatens the TPOT SLO.

\newcommand{\worst}[1]{\cellcolor{red!12}{#1}}
\newcommand{\gain}[1]{{\color{green!50!black}\scriptsize\raisebox{0.3pt}{$\blacktriangledown$}\,#1}}
\newcommand{\loss}[1]{{\color{red!70!black}\scriptsize\raisebox{0.3pt}{$\blacktriangle$}\,#1}}
\newcommand{\eok}[1]{{\color{green!55!black}\scriptsize #1}}
\newcommand{\ebad}[1]{{\color{red!70!black}\scriptsize #1}}
\section{Evaluation}
\label{sec:evaluation}

\subsection{Implementation and Setup}
\label{sec:eval-setup}

\textbf{Implementation.}
We implement KernelFlume (${\sim}$6\,K LoC in Python, C++, and CUDA) on top of vLLM~0.14, reusing its kernels.

\textbf{Hardware and models.}
We evaluate on two real testbeds: (A) an intra-node 8$\times$RTX A6000 testbed (48\,GB GDDR6, ${\sim}$768\,GB/s, hybrid NVLink/PCIe) and (B) a cross-node 16$\times$H100 testbed (two nodes, 8$\times$H100 each, 80\,GB HBM3, ${\sim}$3.35\,TB/s per GPU, NVSwitch intra-node and 400\,Gb/s RoCE with GPUDirect-RDMA inter-node).
Both run Llama-3.1-8B-Instruct with real weights.
For the large-scale simulation (\S\ref{sec:eval-sim}), we simulate Llama-70B, and Llama-3.1-405B for the ultra-large-scale projection, using profiled H100 parameters.

% ==============================================================
% ==============================================================
\subsection{Online Elasticity and Scalability}
\label{sec:eval-controller}

This section evaluates whether adding A~nodes reduces attention latency, and quantifies the topology-switch overhead.

\begin{figure}[t]
    \centering
    \includegraphics[width=\linewidth]{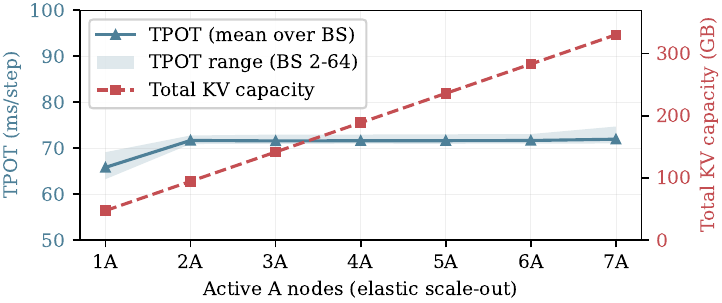}
    \caption{Elastic attention-node scaling from one to seven A~nodes (1A$\to$7A; A6000, Llama-3.1-8B). As nodes are added, total KV capacity grows linearly (red) while per-step TPOT stays flat (blue; mean over batch sizes 2--64, band = min--max spread). A weightless A~node adds capacity with low measured latency overhead.}
    \label{fig:multi-a}
\end{figure}

\paragraph{Attention-node scaling.}
We evaluate whether KernelFlume can add KV capacity without degrading decode latency, by elastically scaling from one attention node (1A) to seven attention nodes (7A) as each request's context grows autoregressively (Figure~\ref{fig:multi-a}): when the tail A~node's KV approaches its budget, the controller appends a new weightless A~node.
We sweep six batch sizes (2--64), filling every A~node to its maximum HBM budget for KV.

As KernelFlume scales from one to seven A~nodes, total KV capacity grows linearly from 47 to 330\,GB while per-step TPOT stays nearly flat (${\sim}$72\,ms).
The narrow min--max band shows only small variation across batch sizes, with TPOT remaining stable as A~nodes are added.
The small rise before the first A-node scale-up is the normal effect of growing sequence length, before elasticity is needed; after scale-up, weightless A~nodes absorb the incremental KV demand while keeping per-step latency nearly flat.

\begin{figure}[t]
    \centering
    \begin{subfigure}[t]{0.27\linewidth}
      \centering
      \includegraphics[width=\linewidth]{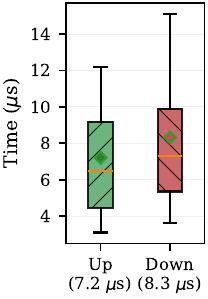}
      \caption{Route-install}
      \label{fig:switch-boxplot}
    \end{subfigure}%
    \hfill
    \begin{subfigure}[t]{0.69\linewidth}
      \centering
      \includegraphics[width=\linewidth]{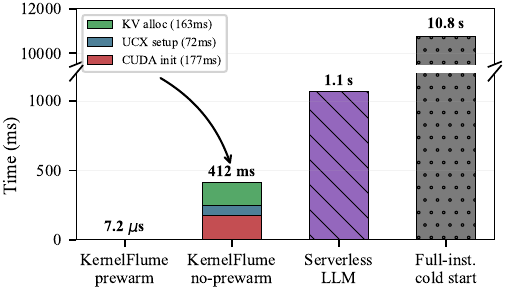}
      \caption{Scaling overhead comparison}
      \label{fig:switch-comparison}
    \end{subfigure}
    \caption{Topology-switch overhead.
    (a)~Prepared route-install overhead distribution (mean scale-up: 7.2\,$\mu$s, scale-down: 8.3\,$\mu$s).
    (b)~Comparison with full-instance scaling: prewarmed route-install takes 7.2\,$\mu$s, and even no-prewarm A-node bring-up (412\,ms, no weights) is shorter than the 1.1\,s ServerlessLLM and 10.8\,s full-instance disk cold start.}
    \label{fig:switch-overhead}
\end{figure}

\paragraph{Topology-switch overhead.}
We measure topology-switch overhead over ${\sim}$30 elasticity events (Figure~\ref{fig:switch-overhead}).
With background pre-warming, activating an A~node only installs a route (\texttt{a\_eps.append}); CUDA context, UCX transport, and KV-cache setup occur off the critical path, and no model weights are loaded.
We compare against two full-instance startup methods: ServerlessLLM~\cite{fu2024serverlessllm}, a state-of-the-art optimized loader, and a full-instance disk cold start.

Prepared route-install takes 7.2\,$\mu$s on scale-up and 8.3\,$\mu$s on scale-down, because activation changes only the routing table: A~node setup is prewarmed and never loads model weights.
The predictive policy (\S\ref{sec:scaling-policy}) normally starts this setup $\tau_{\text{fill}}$ steps before activation. If a burst outruns prewarming, a weightless A~node can still be built from scratch in 412\,ms (CUDA context, UCX init, and KV allocation), shorter than the 1.1\,s ServerlessLLM loader and the 10.8\,s full-instance disk cold start.

% ==============================================================
\subsection{Asynchronous Kernel Execution}
\label{sec:eval-engine}

This section evaluates the two execution mechanisms in \S\ref{sec:pipeline}: query-first attention disaggregation and kernel pipelining.

\paragraph{Disaggregation overhead and query-first attention.}

\begin{figure}[t]
    \centering
    \includegraphics[width=\linewidth]{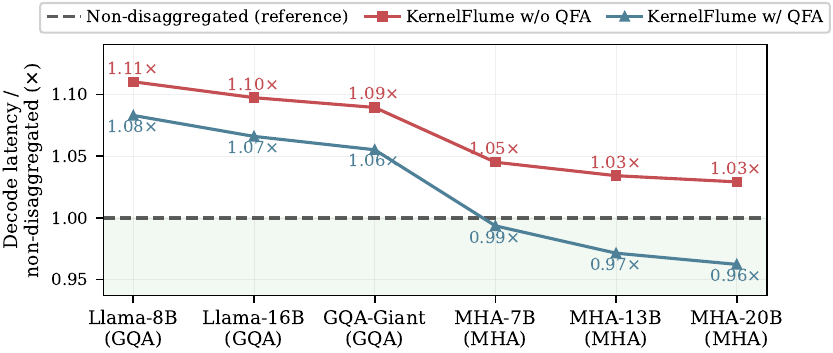}
    \caption{Decode latency of disaggregated execution relative to the non-disaggregated reference (a single GPU, no W--A split; $1.0\times$, lower is better) across six GQA/MHA models. With QFA, disaggregated decode stays within a few percent of the reference, and runs \emph{faster} than it for the MHA models.}
    \Description{Line chart across six models (three GQA, three MHA) of the disaggregated-to-non-disaggregated decode-latency ratio, with a dashed reference line at 1.0. KernelFlume without QFA sits at 1.03--1.11$\times$; KernelFlume with QFA is lower, at 1.06--1.08$\times$ for GQA and 0.96--0.99$\times$ for MHA, below the reference.}
    \label{fig:overlap-efficiency}
\end{figure}
We evaluate the overhead of W--A disaggregation and how much query-first attention (QFA) recovers. To isolate the effect of the query-to-KV-head ratio, we use six representative GQA/MHA configurations that vary query-head and KV-head counts (GQA: Q32:KV8, Q48:KV8, Q64:KV8; MHA: Q32:KV32, Q40:KV40, Q48:KV48), and sweep batch sizes 8--128, sequence lengths 2K--32K, and multiple $N_A$ settings.
For each model we compare three setups: \emph{non-disaggregated reference} (a single GPU running attention and projection together, i.e., a W--A split with zero split overhead), \emph{KernelFlume w/o QFA} (synchronous W--A split), and \emph{KernelFlume w/ QFA} (with query-first overlap enabled).
We report each disaggregated setup's decode latency relative to this reference (measured\,/\,reference latency; $1.0\times$ is on par, lower is better), averaged across all (batch, seq) combinations per model (Figure~\ref{fig:overlap-efficiency}).

Without QFA, disaggregation adds 3\% overhead for the large MHA configurations and up to 11\% for the smallest GQA configuration (Q32:KV8), mainly from per-layer UCX round trips that are better amortized by larger GEMMs.
QFA hides this communication behind local projection, recovering 2--3\% for GQA models and 5--7\% for MHA models (${\approx}4\%$ on average).
The effect is strongest for MHA: its heavier K/V projection gives QFA more local work to overlap with remote attention, so the 5--7\% recovery more than absorbs the split overhead, making disaggregated MHA decode \emph{faster} than the non-disaggregated reference (0.96--0.99$\times$).

\begin{figure}[t]
    \centering
    \begin{subfigure}[t]{0.315\linewidth}
        \centering
        \includegraphics[width=\linewidth]{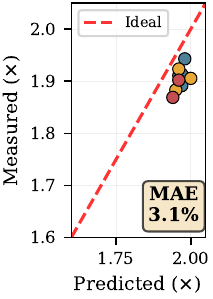}
        \caption{Predicted vs.\ measured}
        \label{fig:pipeline-scatter}
    \end{subfigure}%
    \hfill
    \begin{subfigure}[t]{0.655\linewidth}
        \centering
        \includegraphics[width=\linewidth]{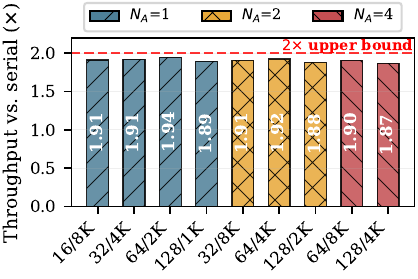}
        \caption{Pipelined\,/\,serial throughput}
        \label{fig:pipeline-bar}
    \end{subfigure}
    \caption{Pipeline-model validation and measured kernel-pipeline throughput ratio across 9 workloads.
    (a) Predicted and measured $M{=}2$ pipelined-to-serial ratios agree with 3.1\% mean absolute error.
    (b) With $N_A^*$ provisioned by the elastic scaling policy (colors), all configurations approach the $2\times$ upper bound.}
    \Description{Two-panel figure. The left panel is a scatter plot of predicted versus measured pipelined-to-serial throughput ratio for 9 workloads, with points near the ideal diagonal and mean absolute error of 3.1 percent. The right panel is a bar chart of measured pipelined-to-serial throughput ratio for the same 9 configurations, colored by the selected number of attention nodes.}
    \label{fig:pipeline-model}
\end{figure}
\paragraph{Kernel pipelining.}
We evaluate kernel pipelining by measuring pipelined throughput divided by serial decode throughput across a range of workloads.
We sample 9 workloads covering batch sizes 16--128 and sequence lengths 2K--16K.
For each workload, the elastic scaling policy chooses $M_{\text{thresh}}$ from the KV and TPOT constraints and sets $N_A^*$ from total KV demand.
The resulting configurations use one, two, or four A~nodes.
We then deploy the corresponding 1W+$N_A^*$A topology and compare the model's predicted throughput ratio against measurement (Figure~\ref{fig:pipeline-model}).

Figure~\ref{fig:pipeline-model} shows that the pipeline model predicts the measured pipelined-to-serial throughput ratio with 3.1\% mean absolute error, supporting its use in the scaling policy.
Pipelining then consistently approaches the $2\times$ upper bound: measured ratios range from 1.87$\times$ to 1.94$\times$ (mean 1.90$\times$), with the residual gap attributable to communication overhead.
This indicates that the $C_{\text{thresh}}$-based provisioning keeps exposed attention-side work within the modeled SLO budget, and kernel pipelining effectively recovers the utilization lost to disaggregation.

\subsection{End-to-End Elasticity}
\label{sec:eval-e2e}

This end-to-end experiment evaluates KernelFlume on the real Codex/SWE-bench Pro agentic workload, against static and full-instance elastic baselines.
We report p99 TPOT, SLO attainment, and cost per million output tokens (\$/Mtok).
SLO attainment is the fraction of decode steps within the per-testbed TPOT target (80\,ms on A6000, 50\,ms on H100); cost is active GPU-hours times the per-GPU hourly price~\cite{cloudgpu2026pricing}, normalized by output tokens.

% --- NEW: merged A6000+H100 end-to-end figure ---
\begin{figure*}[t]
    \centering
    \includegraphics[width=\textwidth]{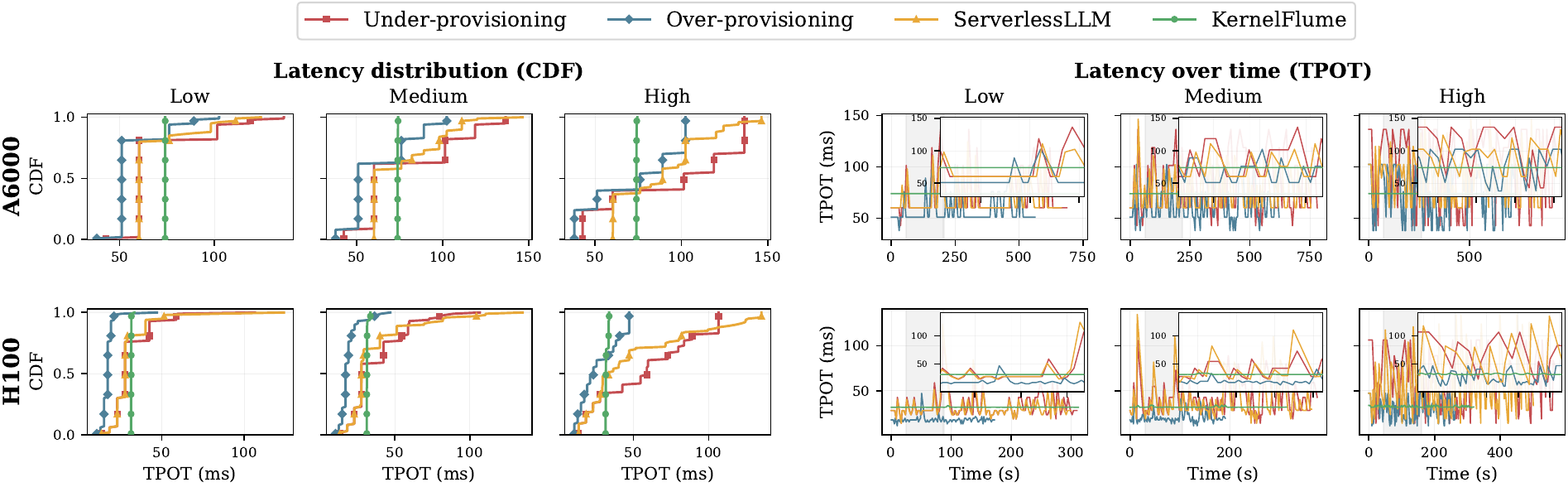}
    \caption{End-to-end elasticity under the Codex/SWE-bench Pro agentic trace on both testbeds.
    Top row: intra-node A6000; bottom row: cross-node 16$\times$H100.
    Within each row, the left panels show TPOT CDFs and the right panels show TPOT over time for the \textsc{Low}, \textsc{Medium}, and \textsc{High} workloads.
    KernelFlume (green) keeps TPOT tight as variance grows; the static and ServerlessLLM baselines develop heavier tails.}
    \Description{A two-by-six grid: the top row is the intra-node A6000 testbed and the bottom row is the cross-node 16$\times$H100 testbed; within each row the left three columns are TPOT CDFs and the right three are per-step TPOT over time, across Low/Medium/High heterogeneity.}
    \label{fig:e2e-merged}
\end{figure*}

We use the open-sourced Codex/SWE-bench Pro serving trace~\cite{qiao2026vllmmooncake,inferact2026codextraces}, which contains 610 tool-augmented coding sessions with long contexts (median roughly 80K tokens, tail to ${\sim}$273K).
To stress the systems under increasing request-length dynamics, we replay three 100-session workloads, \textsc{Low}, \textsc{Medium}, and \textsc{High}, that sample low-, medium-, and high-dynamics slices of the real session-length distribution; \textsc{Low} fluctuates mildly around the trace median while \textsc{High} has the heaviest length spread.
We compare against the baselines from \S\ref{sec:motivation}: fixed-pool under-provisioning, fixed-pool over-provisioning, and full-instance elastic scaling realized with ServerlessLLM~\cite{fu2024serverlessllm}, the state-of-the-art instance-startup method. The baselines use vLLM~\cite{kwon2023vllm}-style decode execution aligned with KernelFlume on decode kernels; they differ in provisioning/scaling policy and DP$\times$CP layout.
We replay the same trace on A6000 at batch size~8 and on a cross-node 16$\times$H100 testbed at batch size~32. In all non-KernelFlume baselines, each serving rank is weight-bearing: HBM is first reserved for model weights, and only the remaining capacity is available for KV. In KernelFlume, only W~nodes store weights, while weightless A~nodes devote their HBM to KV and are appended on demand.

For fixed-pool provisioning, we give each GPU budget its strongest DP$\times$CP capacity envelope by sweeping all factorizations and reporting the best result, using a KV-resident CP implementation consistent with KernelFlume's attention-side CP.
Under- and over-provisioning size the pool for trace-average and trace-maximum KV demand, respectively, with over-provisioning requiring six GPUs on A6000 and fourteen on H100.

Figure~\ref{fig:e2e-merged} and Table~\ref{tab:e2e-summary} summarize the end-to-end results on both testbeds.
KernelFlume is the lowest-cost strategy on all six workloads, cutting cost over ServerlessLLM by 21--32\% (A6000) and 27--61\% (H100) while holding 100\% SLO attainment at flat p99 TPOT (${\sim}$74\,ms and ${\sim}$34\,ms, respectively).
The baselines expose the expected tradeoff: under-provisioning overflows under long contexts, ServerlessLLM pays cold-start latency, and over-provisioning pays for trace-maximum capacity.
KernelFlume avoids this tradeoff in two ways: it removes repeated weight copies when HBM is the binding resource, and it expands dynamic KV capacity with weightless A~nodes without fixed-pool overflow or full-instance cold starts~\cite{fu2024serverlessllm,wu2025dualpath,chen2025concur}.

% ==============================================================
%
%
%
%
%

% ==============================================================

% --- e2e tables placed at the end of the subsection so both float below the e2e figure (Fig.~\ref{fig:e2e-merged}) on the same page ---
\begin{table}[t]
    \caption{End-to-end under the Codex/SWE-bench Pro trace on an intra-node A6000 and a cross-node 16$\times$H100 testbed. Each cell reports p99 TPOT (ms)\,/\,SLO attainment and cost (\$/Mtok), defined in \S\ref{sec:eval-e2e}. Best \$/Mtok bolded; the last column (vs.\ SLLM) is KernelFlume's reduction over ServerlessLLM (SLLM)~\cite{fu2024serverlessllm}, the full-instance baseline.}
    \label{tab:e2e-summary}
    \footnotesize
    \setlength{\tabcolsep}{1.2pt}
    \begin{tabular}{l*{3}{rr}rrr}
    \toprule
     & \multicolumn{4}{c}{Static deployment} & \multicolumn{5}{c}{Elastic scaling} \\
    \cmidrule(lr){2-5}\cmidrule(lr){6-10}
     & \multicolumn{2}{c}{Under-prov.} & \multicolumn{2}{c}{Over-prov.} & \multicolumn{2}{c}{ServerlessLLM} & \multicolumn{3}{c}{KernelFlume} \\
    \cmidrule(lr){2-3}\cmidrule(lr){4-5}\cmidrule(lr){6-7}\cmidrule(lr){8-10}
    Level & \scriptsize p99/slo & \scriptsize \$/Mtok & \scriptsize p99/slo & \scriptsize \$/Mtok & \scriptsize p99/slo & \scriptsize \$/Mtok & \scriptsize p99/slo & \scriptsize \$/Mtok & \scriptsize vs.\ SLLM \\
    \midrule
    \multicolumn{10}{l}{\emph{8$\times$A6000 (48\,GB), intra-node, batch size 8}} \\
    Low      & 136/81 & 4.8 & 102/94 & 5.9 & 124/85 & 5.3 & 74/100 & \textbf{4.2} & \gain{21\%} \\
    Medium   & 136/62 & 5.5 & 102/82 & 6.5 & 146/64 & 6.4 & 74/100 & \textbf{4.5} & \gain{29\%} \\
    High     & 136/40 & 6.5 & 102/54 & 7.6 & 146/42 & 7.8 & 74/100 & \textbf{5.3} & \gain{32\%} \\
    \midrule
    \multicolumn{10}{l}{\emph{16$\times$H100 (80\,GB), cross-node, batch size 32}} \\
    Low     & 62/76 & 5.6 & 30/99 & 7.2 & 85/94 & 5.7 & 32/100 & \textbf{4.1} & \gain{27\%} \\
    Medium  & 89/58 & 6.7 & 59/95 & 8.2 & 126/81 & 8.7 & 34/100 & \textbf{4.8} & \gain{45\%} \\
    High    & 119/34 & 9.6 & 60/67 & 12.5 & 148/55 & 17.7 & 34/100 & \textbf{6.9} & \gain{61\%} \\
    \bottomrule
    \end{tabular}
    \end{table}
% ==============================================================

\subsection{Simulation}
\label{sec:eval-sim}

To study larger models, longer contexts, and heterogeneous hardware, we replay the trace workloads in a simulator using measured hardware parameters. It uses the same metrics and GPU-hour accounting as \S\ref{sec:eval-e2e}.

\paragraph{Validation and 70B-scale simulation.}
We validate the simulator against the measured cross-node 8B H100 setup in \S\ref{sec:eval-e2e}, matching p99 TPOT and cost with 2.8\% mean absolute error (7.5\% max), and then replay the Codex/SWE-bench Pro trace~\cite{qiao2026vllmmooncake,inferact2026codextraces} at Llama-70B scale (tensor parallelism TP=8, batch size BS=128).
The simulator uses standard compute~\cite{williams2009roofline} and communication~\cite{hockney1994communication} models with H100-measured parameters; heterogeneous W/A placement substitutes an H20 profile for A~nodes.
We use an 80\,ms TPOT SLO for the 70B study, preserving comparable headroom over measured decode.

\begin{table}[t]
\caption{Simulation: Llama-70B TP=8, BS=128. Cells report p99 TPOT (ms)\,/\,SLO attainment (\%) and \$/Mtok; best \$/Mtok per row bolded.}
\label{tab:sim-summary}
\scriptsize
\setlength{\tabcolsep}{0.5pt}
\begin{tabular}{l*{3}{rr}rrr}
\toprule
 & \multicolumn{2}{c}{Under-prov.} & \multicolumn{2}{c}{Over-prov.} & \multicolumn{2}{c}{ServerlessLLM} & \multicolumn{3}{c}{KernelFlume} \\
 & \scriptsize p99/slo & \scriptsize \$/Mtok & \scriptsize p99/slo & \scriptsize \$/Mtok & \scriptsize p99/slo & \scriptsize \$/Mtok & \scriptsize p99/slo & \scriptsize \$/Mtok & \scriptsize vs.\ SLLM \\
\midrule
Low      & 171/40 & 28.9 & 41/100 & 56.0 & 175/72 & 28.9 &  49/100 & \textbf{12.6} & \gain{56\%} \\
Medium   & 173/31 & 32.8 & 41/100 & 57.7 & 197/70 & 40.3 &  49/100 & \textbf{14.5} & \gain{64\%} \\
High     & 323/18 & 39.5 & 54/100 & 60.7 & 329/69 & 51.9 &  50/100 & \textbf{17.8} & \gain{66\%} \\
\bottomrule
\end{tabular}

\end{table}

Table~\ref{tab:sim-summary} shows that KernelFlume keeps p99 TPOT at 49--50\,ms with 100\% SLO attainment and reduces cost by 56--66\% over ServerlessLLM.
The 70B simulation reproduces the measured trend: KernelFlume matches over-provisioning's 100\% SLO attainment at 3.4--4.4$\times$ lower cost.
Disaggregation also unlocks a cheaper heterogeneous deployment: because A~nodes are memory-bandwidth-bound, they need not sit on expensive high-compute GPUs. Placing them on bandwidth-rich H20 GPUs while keeping W~nodes on H100, beyond the homogeneous numbers in Table~\ref{tab:sim-summary}, would widen the cost reduction to 80--85\%.

\paragraph{Projection to ultra-large scale.}
We use the simulator to project to ultra-large scale: Llama-3.1-405B in its deployment (FP8 weights on a single 8$\times$H100 node, TP=8) under contexts extended to $1$M--$100$M tokens, far beyond today's trace tail. At this model size the weights dominate every GPU: the 405B FP8 weights occupy $50.6$\,GB of each $80$\,GB card (TP-sharded), leaving only ${\sim}29$\,GB for KV, whereas KernelFlume's weightless A~nodes spend their full ${\sim}78$\,GB on KV, a $2.65\times$ higher KV density per GPU. KernelFlume therefore requires only 30--38\% as many GPUs as ServerlessLLM and cuts cost by 65--72\% across the $1$M--$100$M range, widening to 88--89\% with the same heterogeneous H20 placement.

\section{Related Work}
\label{sec:related}

\textbf{General serving optimizations.}
Existing LLM serving systems improve decode efficiency through scheduling, batching, memory management, and parallel execution, including continuous batching~\cite{yu2022orca}, chunked prefill~\cite{agrawal2024sarathi}, prefix reuse~\cite{zheng2024sglang}, paged attention~\cite{kwon2023vllm}, tensor parallelism~\cite{shoeybi2019megatron}, and MoE expert placement~\cite{zhu2025megascale,liu2024deepseek,deepep2025}.
Other systems extend effective KV capacity through compression or offloading~\cite{sheng2023flexgen}.
These techniques are complementary to KernelFlume: they improve execution within a fixed deployment or move KV along the memory hierarchy, whereas KernelFlume changes the marginal scaling unit for decode-time KV capacity.

\textbf{Prefill/decode disaggregation.}
DistServe~\cite{zhong2024distserve} and Splitwise~\cite{patel2024splitwise} separate the compute-bound prefill phase from the memory-bound decode phase onto specialized hardware, significantly improving GPU utilization. Mooncake~\cite{qin2024mooncake} extends this idea with a KVCache-centric architecture for cross-datacenter serving. KernelFlume inherits this disaggregation philosophy but pushes it to a finer granularity: it separates two kernel classes (projection/FFN and attention) \emph{within} every decode step, rather than two phases that run sequentially. This intra-decode split exposes an elastic dimension that phase-level PD systems leave inside the decode unit.

\textbf{Disaggregated KV and remote attention.}
Distributed-KV systems are related to KernelFlume in how they decouple KV capacity from a single model instance. MemServe~\cite{hu2024memserve} exposes an elastic memory pool that manages and transfers KV blocks across serving instances, and Infinite-LLM~\cite{lin2024infinitellm} partitions KV blocks and can compute attention near the remote KV; both establish that KV \emph{placement} can be decoupled from a single model instance. KernelFlume differs in the \emph{scaling unit}: it adds \emph{weightless} KV-resident attention nodes that store KV and compute attention locally, returning compact partial-attention results $(o,m,\ell)$, so aggregate KV/HBM capacity grows without replicating projection/FFN weights on the unit that scales.

\textbf{Context-parallel reconfiguration.}
LoongServe~\cite{bai2024loongserve} addresses online reconfiguration. Unlike static context parallelism~\cite{yang2024context,liu2023ring}, it forms sequence-parallel groups from a fixed pool of weight-bearing instances, with KV placement and query dispatch. KernelFlume shares query dispatch to nodes that hold KV, but targets a different elastic unit: LoongServe improves packing within a fixed pool of weight-bearing ranks, whereas KernelFlume appends weightless A~nodes to expand physical KV capacity.

\section{Discussion}
\label{sec:discussion}
KernelFlume targets decode-time elasticity, where KV state grows while model weights remain fixed.
Scale-up appends token ranges to A~nodes through routing-table updates, instead of rebuilding collective groups or launching another model replica.
This focus leaves prefill-side optimizations independent: prefix caching, chunked prefill, and prefill/decode disaggregation can be used alongside KernelFlume.

Under the scaling policy in \S\ref{sec:scaling-policy}, KernelFlume assigns newly generated token ranges to newly added A~nodes and releases a retiring A~node once the requests holding its ranges complete.
End-to-end cost is computed from active GPU-hours rather than a static capacity label (\S\ref{sec:eval-e2e}).
A W~node may itself be a tensor-parallel group for large models.
In that case, TP all-reduces remain inside the W~group, while A-node scale events only update attention routes.
Thus, adding or removing A~nodes does not change the W-side TP group.
KernelFlume currently partitions KV by sequence ranges, matching the dominant growth dimension in long-context decode.
Production deployments can further add KV replication, failure recovery, and migration-based compaction on top of the same routing-table abstraction.
Extending the routing policy to hybrid sequence, batch, and head partitioning is a natural direction for future work.

\section{Conclusion}

KernelFlume is built on a simple claim: in decode-heavy long-context serving, fixed model weights and growing per-request KV state should not share the same elastic unit.
It keeps model weights on weight nodes, scales weightless attention nodes with request-state demand, and makes this split practical through step-boundary route updates, query-first attention disaggregation, and inter-layer kernel pipelining.
On a real intra-node A6000 testbed and a cross-node H100 testbed under a real-world dynamic long-context agentic workload, KernelFlume delivers a flat p99 TPOT with up to 32\% (A6000) and 61\% (H100) lower cost per million output tokens than full-instance elastic scaling with ServerlessLLM.
Beyond the measured testbeds, replaying the same trace at Llama-70B scale with measured hardware parameters projects a 56--66\% cost reduction over ServerlessLLM, widening to 80--85\% with heterogeneous attention-node hardware, with the advantage persisting into the million-token context range.
These results show that aligning elasticity with KV demand turns capacity changes from heavyweight instance management into lightweight route updates, improving both tail latency and serving cost.

%% No acknowledgments in this version.

\bibliographystyle{ACM-Reference-Format}
\bibliography{sample-base}

% \appendix
% TODO: Add appendix content if needed.

\end{document}